 \documentstyle[nato,epsfig]{crckapb} 

\newcommand{\ba}{\begin{eqnarray}}
\newcommand{\ea}{\end{eqnarray}}
\newcommand{\bmath}{\begin{mathletters}}
\newcommand{\emath}{\end{mathletters}}
\newcommand{\ban}{\begin{eqnarray*}}
\newcommand{\ean}{\end{eqnarray*}}
\newcommand{\tl}{\tilde{\ell}}

\begin{opening}
\title{RELATIVISTIC PSEUDOSPIN SYMMETRY in NUCLEI}

\author{J.N. Ginocchio$^1$ and A. Leviatan$^2$} 
\institute{$^{1}$Theoretical Division, Los Alamos National Laboratory,\\ 
Los Alamos, New Mexico 87545, USA\protect\\
$^{2}$Racah Institute of Physics, The Hebrew University,\\
Jerusalem 91904, Israel}
\end{opening}

\runningtitle{RELATIVISTIC PSEUDOSPIN SYMMETRY in NUCLEI}

\begin{document}


\section{Introduction}

Peter Ring was one of the first to really grasp the significance of 
pseudospin symmetry as a relativistic symmetry \cite{gino,ami,ginomad,ring}. 
Originally, pseudospin doublets were introduced into nuclear physics 
to accommodate an observed near degeneracy of certain normal-parity 
shell-model orbitals with non-relativistic quantum numbers
($n_r$, $\ell$, $j = \ell + 1/2)$ and
($n_{r}-1, \ell + 2$, $j = \ell + 3/2$)
where $n_r$, $\ell$, and $j$ are the
single-nucleon radial, orbital, and total  angular momentum quantum
numbers, respectively \cite{aa,kth}. The doublet structure, 
is expressed in terms of a ``pseudo'' orbital angular momentum
$\tilde{\ell}$ = $\ell$ + 1 coupled to a ``pseudo'' spin, $\tilde s$ = 1/2. 
For example, $(n_r s_{1/2},(n_r-1) d_{3/2})$ will have
$\tilde{\ell}= 1$, $(n_r p_{3/2},(n_r-1) f_{5/2})$ will have
$\tilde{\ell}= 2$, etc. Since $j = \tilde{\ell}\pm \tilde s$, 
the energy of the two states
in the doublet is then approximately independent of the orientation
of the pseudospin. Some examples are given in Table 1.
In the presence of deformation the doublets 
persist with asymptotic (Nilsson) quantum numbers 
$[N,n_3,\Lambda,\Omega=\Lambda+1/2]$ and 
$[N,n_3,\Lambda+2,\Omega=\Lambda+3/2]$, and can be expressed in terms 
of pseudo-orbital and total angular momentum projections 
$\tilde{\Lambda}=\Lambda+1$, $\Omega=\tilde\Lambda\pm 1/2$.
This pseudospin ``symmetry''
has been used to explain features of deformed nuclei \cite{bohr},
including superdeformation \cite{dudek}
and identical bands \cite{twin,stephens,von}. 
While pseudospin symmetry is experimentally well corroborated in nuclei, 
its foundations remained a mystery and ``no deeper understanding of
the  origin of these (approximate) degeneracies'' existed
\cite {ben}. In this contribution we review more recent
developments that show that pseudospin symmetry is an approximate 
relativistic symmetry of the Dirac Hamiltonian with realistic nuclear 
mean field potentials \cite{gino,ami}.

\begin{table}[t]
\begin{center}
\caption{Experimental (Exp) and relativistic mean field (RMF) 
pseudospin binding energy splittings $\epsilon_{j^{\prime}=
\tilde{\ell}+1/2}-\epsilon_{j= \tilde{\ell}-1/2}$ for various doublets in
$^{208}$Pb.}

\begin{tabular}{|c|c|c|c|}
\hline

    $\tilde{\ell}$ & ($n_{r}-1, \ell + 2$, $j^{\prime} = \ell + 3/2$)
&
$\epsilon_{j^{\prime}= \tilde{\ell}+1/2}-\epsilon_{j= \tilde{\ell}-1/2}$ &
$\epsilon_{j^{\prime}=
\tilde{\ell}+1/2}-\epsilon_{j= \tilde{\ell}-1/2}$  \\
          &-($n_r$,
$\ell$, $j =
\ell + 1/2)$    & (Exp) & (RMF) \protect\cite{ginomad} \\
         &    & (MeV) & (MeV)  \\
\hline

   4 &  $0h_{9/2}-1f_{7/2}$  &  1.073 &    2.575   \\
3 &  $0g_{7/2}-1d_{5/2}$  &  1.791   &  4.333   \\
    2 &  $1f_{5/2}-2p_{3/2}$  &-0.328  &   0.697 \\
1 &  $1d_{3/2}-2s_{1/2}$  &  0.351   &  1.247\\
\hline
\end{tabular}
\end{center}
\end{table}

\section{Pseudospin Symmetry of the Dirac Hamiltonian}

The Dirac Hamiltonian, $H$,
with an external scalar, $V_S$, and vector, $V_V$, potentials is 
invariant under a SU(2) algebra for $V_S = -V_V$ 
leading to pseudospin symmetry in nuclei \cite{ami}. 
The pseudospin generators, $\hat{\tilde{S}}_{\mu}$,
which satisfy $[\,H\,,\, \hat{\tilde{S}}_{\mu}\,] = 0$ in the symmetry 
limit, are given by
\ba
{\hat{\tilde {S}}}_{\mu} =
\left (
\begin{array}{cc}
\hat {\tilde s}_{\mu} &  0 \\
0 & {\hat s}_{\mu}
\end{array}
\right )
= \left (
\begin{array}{cc}
U_p\, {\hat s}_{\mu} \, U_p & 0 \\
0 & {\hat s}_{\mu}
\end{array}
\right )
\label{psgen}
\ea
where
${\hat s}_{\mu} = \sigma_{\mu}/2$ are the usual spin generators,
$\sigma_{\mu}$ the Pauli matrices, and
$U_p = \, {\mbox{\boldmath $\sigma\cdot p$} \over p}$ is the
momentum-helicity unitary operator introduced in \cite {draayer}. If,
in addition, the potentials are spherically symmetric, 
the Dirac Hamiltonian has an additional invariant
SU(2) algebra, $[\,H\,,\, \hat{\tilde{L}}_{\mu}\,] = 0$, with the
pseudo-orbital angular
momentum operators given by
$\hat{\tilde{L}}_{\mu} =
\left ( {\hat {\tilde \ell_{\mu}} \atop 0 }
{ 0 \atop { {\hat \ell}_{\mu}} }\right )$, where 
$\hat {\tilde \ell}_{\mu} = U_p\, {\hat \ell}_{\mu}$ $U_p$,
${\hat\ell}_{\mu}= \mbox{ \boldmath $r$}\times \mbox{ \boldmath $p$}$. 
The pseudo spin $\tilde s$ and pseudo orbital angular momentum 
$\tilde \ell$ are seen to be the ordinary spin and orbital angular 
momentum respectively of the lower component of the Dirac wave function. 
The upper components of the two states in the doublet have orbital 
angular momentum $\tilde \ell \pm 1$ for $j=\tilde\ell \pm 1/2$ in 
agreement with the spherical pseudospin doublets originally observed. 
The corresponding radial quantum numbers are discussed in 
\cite{levgino2,proc1}. 
In the pseudospin symmetry limit the two states in the doublet 
$j = {\tilde {\ell}} \pm 1/2$ are degenerate, 
and are connected by the pseudospin generators
$\hat{\tilde{S}}_{\mu}$ of Eq.~(\ref{psgen}).
This implies relationships between the doublet wave functions. 
In particular, since ${\hat {\tilde S}}_{\mu}$ 
have the spin operator ${\hat {s}}_{\mu}$ operating on the lower component
of the Dirac wave function, it follows that the spatial part of these
components will be equal for the two states in the 
doublet within an overall phase, as can be seen in Fig.~1.
\noindent
\begin{figure}
\begin{minipage}{0.48\linewidth}
\epsfig{file=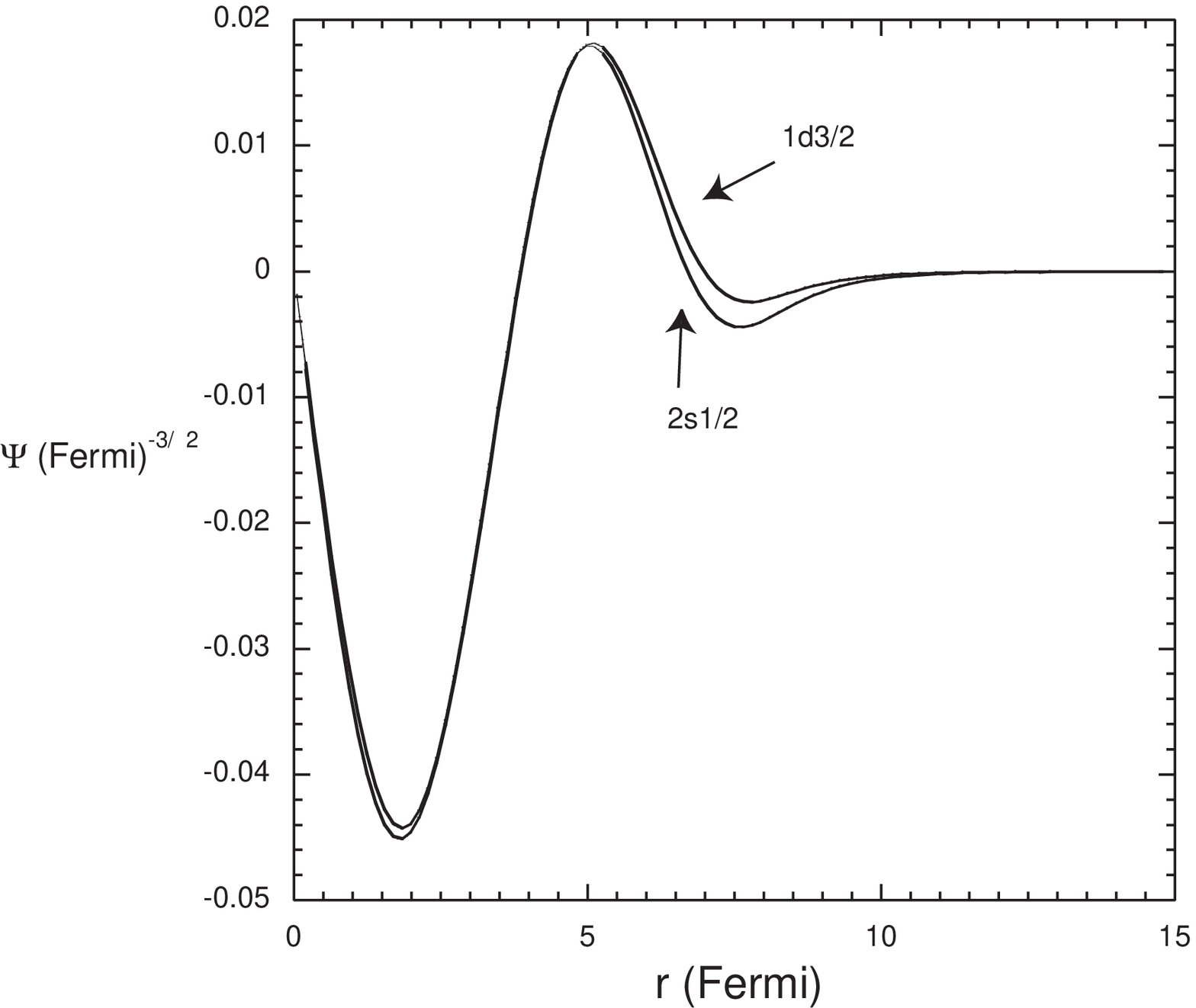,height=6cm,width=\linewidth,angle=0}
\vspace{-0.2cm}
\caption{The lower components of Dirac eigenfunctions 
$(2s_{1/2},1d_{3/2})$ in $^{208}$Pb \protect\cite{ginomad}.}
\end{minipage}
\hspace{\fill}
\begin{minipage}{0.48\linewidth}
\vspace{0.4cm}
\hspace{-4.2cm}
\epsfig{file=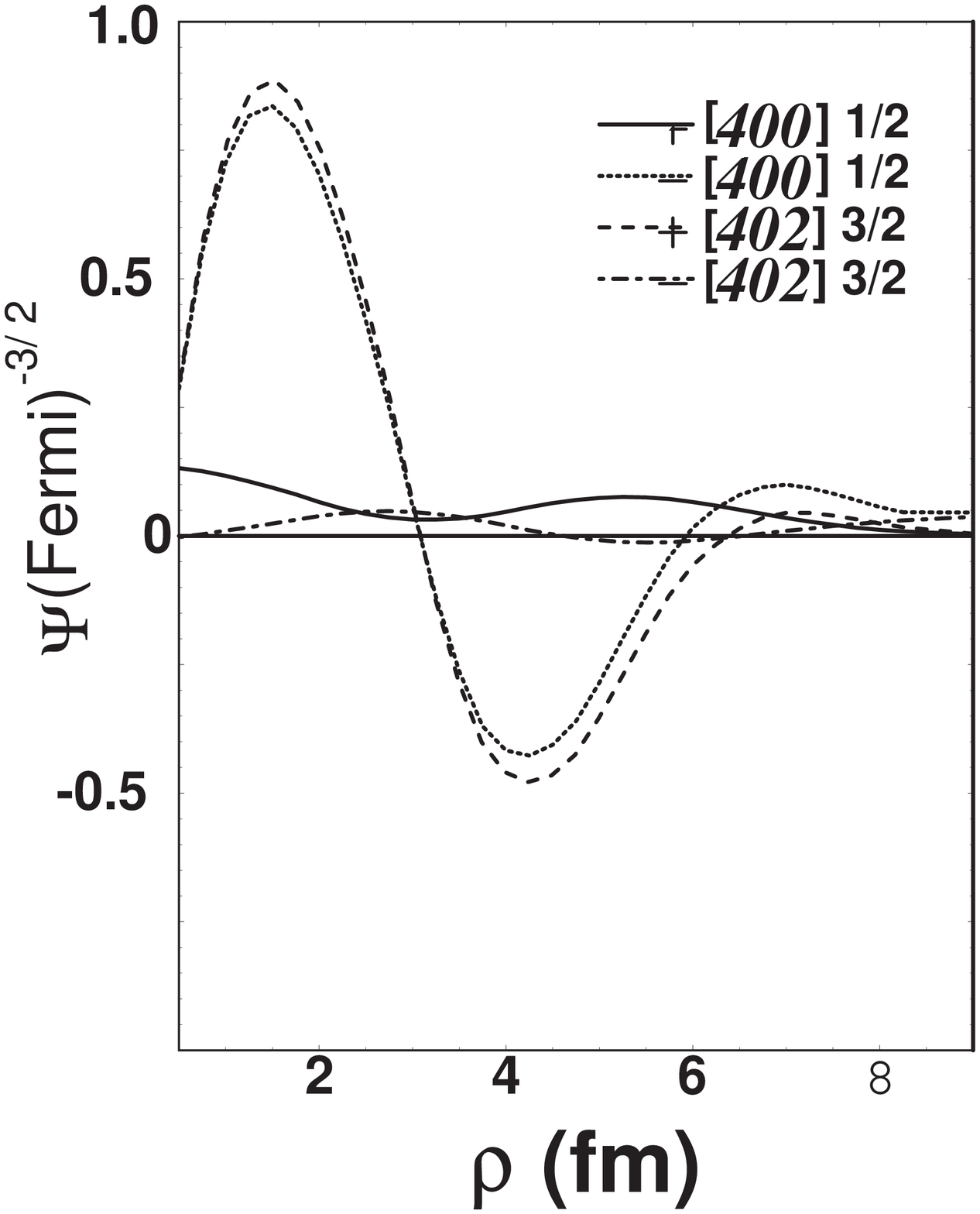,height=7cm,width=10cm,angle=0}
\vspace{-0.4cm}
\caption{ The lower components of Dirac eigenfunctions 
$[400]1/2$ ($+$ solid line, $-$~short-dash line) and 
$[402]3/2$ ($+$ dash line, $-$~dash-dot line) at $z = 1$ 
fm \protect\cite{arima}.}
\end{minipage}
\end{figure}

For axially deformed potentials satisfying $V_S=-V_V$, there is, 
in addition to pseudospin, a conserved U(1) generator, 
corresponding to the pseudo-orbital angular momentum
projection along the body-fixed symmetry axis, 
${\hat{\tilde \lambda}} = \left ( {{\hat {\tilde
\Lambda}} \atop 0 } { 0 \atop {{\hat \Lambda}} }
\right ),$ where ${\hat {\tilde \Lambda}} = U_p\ \hat \Lambda\ U_p$.
In this case the Dirac wave functions are eigenstates of 
${\hat{\tilde \lambda}}$ and both components have the same total angular 
momentum projection $\Omega$. The lower component has 
pseudo-orbital angular momentum projection ${\tilde \Lambda}$ while 
the upper component has ${\tilde \Lambda}\pm 1$ for 
$\Omega = {\tilde \Lambda}\pm 1/2$, in 
agreement with the deformed pseudospin doublets mentioned in Section~1.
For axially deformed nuclei the eigenfunctions depend on two spatial
variables, $z$ and $\rho = \sqrt{ x^2 + y^2}$, and there
are two upper $g_{\pm}(\rho,z)\ \chi_{\pm {1\over 2}}$ and lower
$f_{\pm}(\rho,z)\ \chi_{\pm { 1\over 2}}$
components where $\chi_{\pm { 1\over 2}}$ is 
the spin wavefunction. 
Pseudospin symmetry predicts that, for the pseudospin eigenfunction with
pseudospin projection ${ 1\over 2}$ $(-{ 1\over 2})$, the
lower component $f_{-}(\rho,z)$ $[f_{+}(\rho,z)]$ is zero. 
In addition, the lower component $f_{+}(\rho,z) $ for the pseudospin 
eigenfunction with pseudospin projection ${1\over 2} $ is equal to the the
lower component $f_{-}(\rho,z) $ for the pseudospin eigenfunction 
with pseudospin projection $-{1\over 2}$ up to an overall phase. 
These relations are illustrated in Fig.~2. 

However, in the exact pseudospin limit, $V_S=-V_V$, there are no bound
Dirac valence states. For nuclei to exist the pseudospin symmetry must
therefore be broken. Nevertheless, realistic mean fields involve
an attractive scalar potential and a repulsive vector potential of
nearly equal magnitudes, $V_S \sim - V_V$, and recent 
calculations in a variety of nuclei confirm the existence of an 
approximate pseudospin symmetry in both the energy spectra and 
wave functions \cite{ginomad,ring,arima,ginolev2}.
In Table~1 pseudospin-orbit splittings calculated in the RMF 
\cite {ginomad} are compared with the measured values in the spherical 
nucleus $^{208}$Pb and are seen to be larger than the
measured splittings which demonstrates that the pseudospin
symmetry is better conserved experimentally than mean field theory
would suggest. Figures~1 and 2 show that in realistic RMF calculations, 
the expected relations between the 
lower components of the two states in the doublets are approximately 
satisfied both for spherical and axially deformed nuclei. The behavior 
of the corresponding upper components which dominate the Dirac eigenstates 
is discussed in the next section.

\section{Test of Nuclear Wave Functions for Pseudospin Symmetry}

Since pseudospin symmetry is broken, the
pseudospin partner produced by the raising and lowering
operators acting on an eigenstate will not necessarily be an eigenstate.
The question is how different is the pseudospin
partner from the eigenstate with the same quantum numbers? 
A recent study \cite{ginolev2} addressing this question has 
shown that the radial
wave functions of the upper components of the
$j =  {\tilde {\ell}} - 1/2$ pseudospin partner of the eigenstate with
$j =  {\tilde {\ell}} + 1/2$ is similar in shape to the
$j =  {\tilde {\ell}} - 1/2$ eigenstate but there is a difference in
magnitude. This is shown in Fig.~3 where we compare the $s_{1/2}$ 
pseudospin partners (denoted by $[P(0d_{3/2})]s_{1/2},\; 
[P(1d_{3/2})]s_{1/2}$) of the Dirac eigenstates $\ 0d_{3/2},\ 1d_{3/2}$ 
($\tl=1,\; j=3/2$), for $^{208}$Pb \cite {ginomad}. 
The lower components agree very well, which was noted previously, 
except for some disagreement on the surface. 
For the upper components the agreement is not as good in the magnitude 
but the shapes agree well, with the number of radial nodes being the same.
The agreement improves as the radial quantum number increases, 
which is consistent with 
the observed decrease in the energy splitting between the doublets 
\cite{gino,ginomad}. 
\noindent
\begin{figure}
\vspace{-0.5cm}
\hspace{-0.3cm}
\epsfig{file=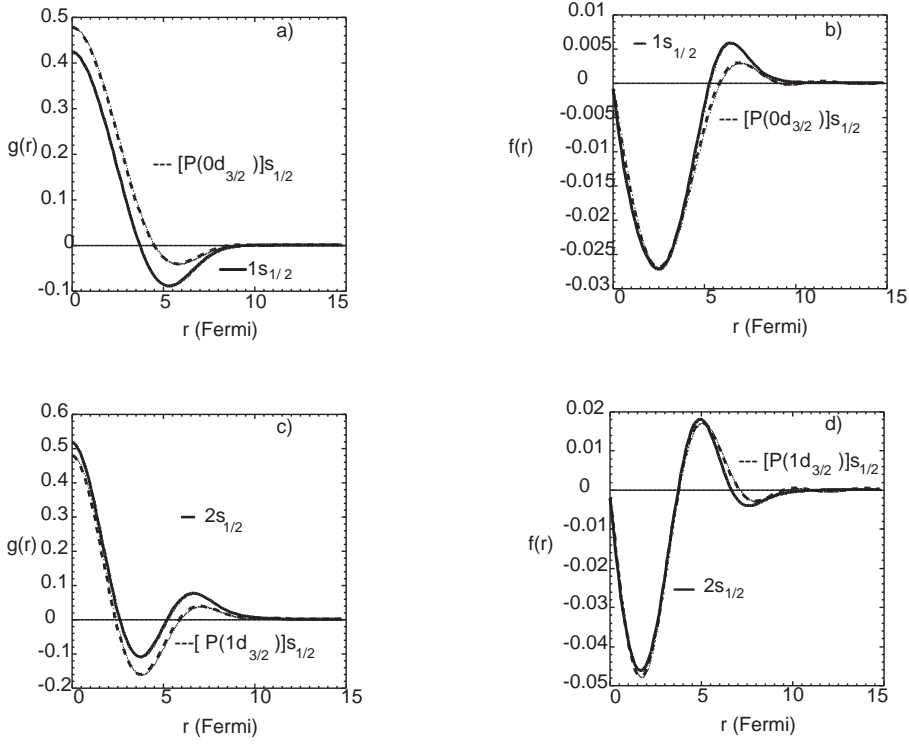,height=10.7cm,angle=0}
\caption{
a) The upper component [$g(r)$] and
b) the lower component [$f(r)]$ in (Fermi)$^{-3/2}$ of the
$[P(0d_{3/2})]s_{1/2}$ partner of the $0d_{3/2}$ eigenstate compared to
the
$1s_{1/2}$ eigenstate.
c)~The upper component and d) the lower component of the
$[P(1d_{3/2})]s_{1/2}$ partner of
the $1d_{3/2}$ eigenstate compared to the $2s_{1/2}$ 
eigenstate for $^{208}$Pb
\protect\cite{ginomad,ginolev2}.}
\end{figure}

On the other hand, the radial wave functions of the upper components of 
the $j =  {\tilde {\ell}} + 1/2$ pseudospin partner of 
the eigenstate with $j =  {\tilde {\ell}} - 1/2$ is not similar in shape
to the $j =  {\tilde {\ell}} + 1/2$ eigenstate. In fact these
wave functions approach $r^{\tilde {\ell}-1}$ rather than
$r^{\tilde {\ell}+1}$ for $r$ small, do not have the same number of
radial nodes as the eigenstates, and do not fall off exponentially as do
the eigenstates, but rather fall off as
$r^{-({{\tilde {\ell}}+2})}$. Furthermore, the pseudospin partners of the
``intruder" nodeless eigenstates, also fall off as as
$r^{-({{\tilde {\ell}}+2})}$. As an example of this category we show in
Fig.~4a,b the radial wavefunction of the $[P(0f_{7/2})]h_{9/2}$ partner
of the $0f_{7/2}$ intruder state ($\tl=4,\;j=7/2$).
There is no quasi-degenerate $h_{9/2}$ eigenstate
to compare to. The upper component has the $r^{-6}$ falloff alluded to above.
Although both components have zero radial quantum number, they do not compare
well with the $0h_{9/2}$ eigenstate shown in Fig. 4c,d. In Fig.~4c,d we
show also the radial wavefunction of the $[P(1f_{7/2})]h_{9/2}$ partner
of the $1f_{7/2}$ state ($\tl=4,\;j=7/2$)
and compare it to the $0h_{9/2}$ eigenstate.
The upper component has again the $r^{-6}$ falloff and therefore does
not compare well on the surface. Also the number of radial quantum numbers
differ. The lower components agree better. 
Work is in progress to study the goodness of pseudospin symmetry in the 
upper components of Dirac eigenfunctions in deformed nuclei.
\noindent
\begin{figure}
\vspace{-0.5cm}
\hspace{0.25cm}
\epsfig{file=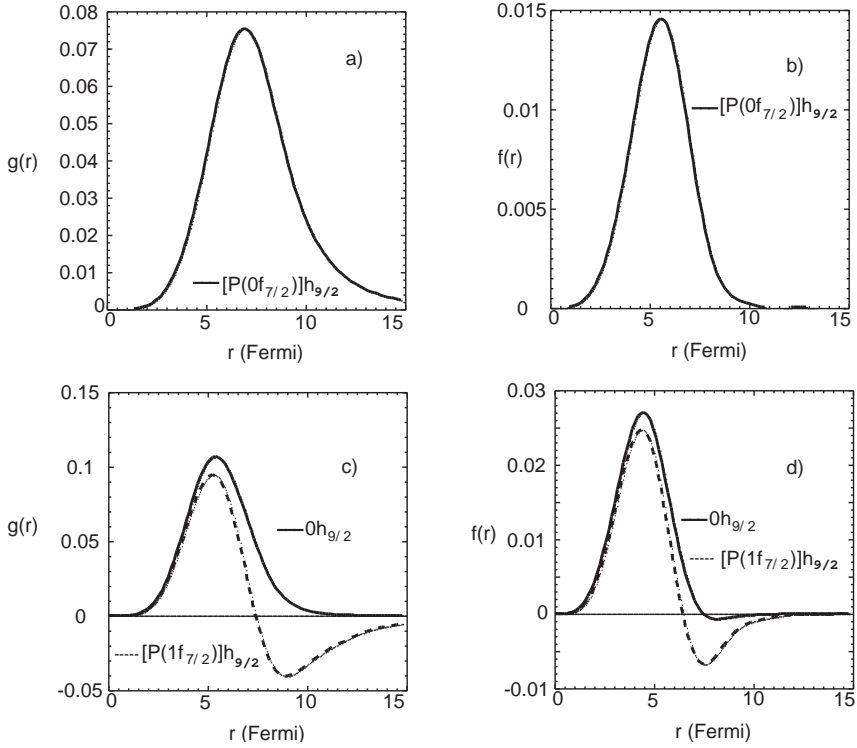,height=11cm,angle=0}
\caption{
a) The upper component [$g(r)$] and
b) the lower component [$f(r)$] in (Fermi)$^{-3/2}$ of the
$[P(0f_{7/2})]h_{9/2}$ partner of the $0f_{7/2}$ eigenstate.
c) The upper component and d) the lower component of the
$[P(1f_{7/2})]h_{9/2}$ partner of the $1f_{7/2}$ eigenstate compared to
the $0h_{9/2}$ eigenstate for $^{208}$Pb \protect\cite{ginomad,ginolev2}.}
\end{figure}

\section{Implications for M1 and Gamow Teller Transitions}

Because the lower components are small, in order to test the pseudospin
symmetry prediction that the lower components are almost identical
we must observe
transitions for which the upper components are not dominant. Magnetic
dipole and Gamow-Teller transitions between the states in the
doublet are forbidden non-relativistically since the orbital angular
momentum of the two states differ by two units, but are allowed
relativistically.
Pseudospin symmetry predicts that, if the magnetic moments, $\mu$, of the two
states are known, the magnetic dipole transition, $B(M1)$, between the states
can be
predicted. Likewise if the Gamow - Teller transitions between the
states with the same quantum numbers are known, the transition
between the states
with different quantum numbers can be predicted \cite {gino3}. For
example for neutrons, the M1 transition is given by
\ba
&&\sqrt{B(M1:{n_{r}-1},{\ell}+2, j^{\prime} \rightarrow {
{n}_{r}}, {\ell}, j} )_{\nu} = 
- \sqrt{j + 1\over 2j + 1}( \mu_{j,\nu} - \mu_{A,\nu}) \qquad
\nonumber\\
&& \qquad\qquad\qquad\qquad
 = {j + 2 \over 2j + 3}\  \sqrt{2j + 1\over j + 1}(\mu_{j^{\prime},\nu} + 
{{j+1}\over {j + 2}} \mu_{A,\nu}) ~,
\label {neutron2}
\ea
where $j^{\prime} = {{\ell}} + 3/2, j = {{\ell}} + 1/2$ and $
\mu_{A,\nu} = -1.913 \mu_{0}$ is the anomalous magnetic moment. A
survey of forbidden
magnetic dipole transitions taking into account the single-particle
corrections by using spectroscopic factors shows a reasonable agreement
with these
relations, an example of which is given in Fig.~5 \cite{peter}. 

\noindent
\begin{figure}
\vspace{-0.09cm}
\hspace{1.3cm}
\epsfig{file=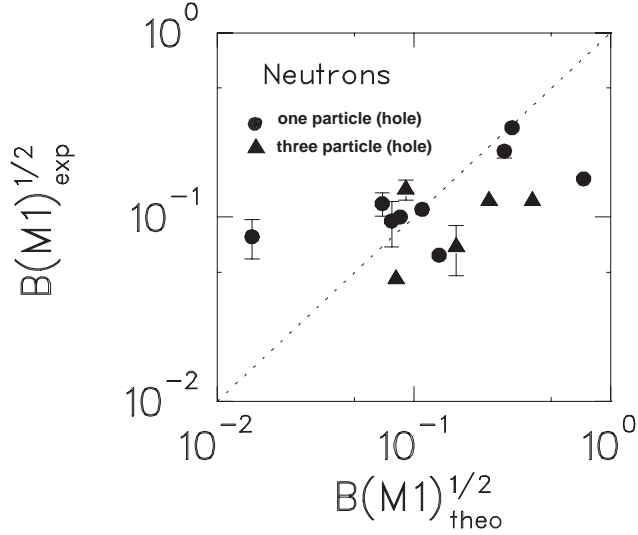,height=7cm,angle=0}
\caption{The experimental and theoretical \protect\cite{peter} 
B(M1) between one-particle or one-hole (circles) states and
three-particle or three hole (triangles) states in the doublet. The
dotted line denotes perfect agreement.}
\end{figure}

\section{Possible Connection to Chiral Symmetry Breaking in QCD}

Applying QCD sum rules in nuclear matter \cite {furn}, the ratio of
the scalar and vector self-energies were determined to be
${V_S \over V_V} \approx - {\sigma_N \over 8 m_q}$ where $\sigma_N $
is the sigma term which can be measured in pion-nucleon scattering
and $m_q$ is the
current quark mass.  For reasonable values of
$\sigma_N
$  and quark masses, this ratio is close to  -1. The significance of
the sigma term is that
it is a measure of chiral symmetry breaking and vanishes if chiral
symmetry is conserved.
The implication
of  these results is
that chiral symmetry breaking is responsible for the scalar field
being approximately equal in magnitude to the vector field, thereby
producing pseudospin symmetry.

\section{Future Outlook}

We have reviewed the foundations and implications of the 
relativistic pseudospin symmetry in nuclei. 
Although not discussed here, pseudospin symmetry
has been shown to be approximately conserved in medium energy nucleon-nucleus
scattering from nuclei \cite {gino4,leeb} but badly broken in low energy
nucleon-nucleus scattering \cite {fred}. 
A recent study \cite{brazil} suggests that pseudospin symmetry will 
improve for neutron rich nuclei. This needs
further investigation. Finally, the link between pseudospin symmetry and
chiral symmetry  breaking in nuclei
needs to be explored at a fundamental level.

\section{Acknowledgements}

D. G. Madland and P. von Neumann Cosel
collaborated on different aspects of the work reported in this survey.
This research was supported in part by the United States Department of
Energy under contract W-7405-ENG-36 and by the U.S.-Israel
Binational Science Foundation.


\begin{thebibliography}{99}

\bibitem{gino}
J.N. Ginocchio, 
Phys. Rev. Lett. {\bf 78} (1997) 436.

\bibitem{ami}
J.N. Ginocchio and  A. Leviatan,
Phys. Lett. B {\bf 425} (1998) 1.

\bibitem{ginomad}
J.N. Ginocchio and D. G. Madland,
Phys. Rev. C {\bf 57} (1998) 1167.

\bibitem{ring}
G.A. Lalazissis, Y.K. Gambhir, J.P. Maharana, C.S. Warke and P. Ring,
Phys. Rev. C {\bf 58} (1998) R45.

\bibitem{aa}
A. Arima, M. Harvey and K. Shimizu,
Phys. Lett. B {\bf 30} (1969) 517.

\bibitem{kth}
K.T. Hecht and A. Adler,
Nucl. Phys. A {\bf 137} (1969) 129.

\bibitem{bohr}
A. Bohr, I. Hamamoto and B.R. Mottelson,
Phys. Scr. {\bf 26} (1982) 267.

\bibitem{dudek}
J. Dudek {\it et al.}, 
Phys. Rev. Lett. {\bf 59} (1987) 1405.

\bibitem{twin}
W. Nazarewicz {\it et al.}, 
Phys. Rev. Lett. {\bf64} (1990) 1654.

\bibitem{stephens}
F.S. Stephens {\it et al.},
Phys. Rev. Lett. {\bf 65} (1990) 301; 
Phys. Rev. C {\bf 57} (1998) R1565.

\bibitem{von}
A.M. Bruce {\it et. al.}, Phys. Rev. C {\bf 56} (1997) 1438.

\bibitem{ben}
B. Mottelson, Nucl. Phys. A {\bf 522} (1991) 1.

\bibitem{draayer}
A. L. Blokhin, C. Bahri, and J.P. Draayer,
Phys. Rev. Lett. {\bf 74} (1995) 4149.

\bibitem{levgino2}
A. Leviatan and J.N. Ginocchio, 
Phys. Lett. B in press, nucl-th/0108016.

\bibitem{proc1}
A. Leviatan and J.N. Ginocchio, contribution to these proceedings.

\bibitem{arima}
J. Meng {\it et al.}, 
Phys. Rev. C {\bf 58} (1998) R628. 

\bibitem{ginolev2}
J.N. Ginocchio and A. Leviatan, 
Phys. Rev. Lett. {\bf 87} (2001) 072502.

\bibitem {gino3}
J.N. Ginocchio, 
Phys. Rev. C {\bf 59} (1999) 2487.

\bibitem {peter}
P. von Neumann Cosel and J.N. Ginocchio, 
Phys. Rev. C {\bf 62} (2000) 014308.

\bibitem{furn}
T.D. Cohen {\it et al.}, 
{\it Prog. in Part. and Nucl. Phys.} {\bf 35} (1995) 221.

\bibitem{gino4}
J. N. Ginocchio, 
Phys. Rev. Lett. {\bf 82} (1999) 4599.

\bibitem {leeb}
H. Leeb and S. Wilmsen, 
Phys. Rev. C {\bf 62} (2000) 024602.

\bibitem{fred}
J. B. Bowlin, A. S. Goldhaber and C. Wilkin, 
Z. Phys. A {\bf 331} (1988) 83.

\bibitem{brazil}
P. Alberto {\it et al.} 
Phys. Rev. Lett. {\bf 86} (2001) 5015.

\end{thebibliography}
\end{document}